\documentclass[aps,prb,preprint,preprintnumbers,amsmath,amssymb,groupedaddress]{revtex4}
\usepackage{amsfonts,amsmath,amssymb}
\usepackage{graphicx}

\begin{document}

\newlength{\figwidth}
\setlength{\figwidth}{0.8\textwidth}


\title{State-to-state rotational transitions in H$_2$+H$_2$ collisions at low temperatures}

\author{Teck-Ghee~Lee}
\affiliation{Department of Physics and Astronomy, University of
Kentucky, Lexington, KY 40506 \\ and Physics Division, Oak Ridge
National Laboratory, Oak Ridge, TN 37831}

\author{N.~Balakrishnan}
 \affiliation{Department of Chemistry, University of Nevada Las Vegas, Las Vegas, Nevada 89154 }

\author{R.~C.~Forrey}
 \affiliation{Department of Physics, Penn State University, Berks-Lehigh Valley College, Reading, PA 19610}

\author{P.~C.~Stancil}
 \affiliation{Department of Physics and Astronomy and Center for Simulational Physics,
 University of Georgia, Athens, GA 30602}

\author{D.~R.~Schultz}
 \affiliation{Physics Division, Oak Ridge National Laboratory, Oak Ridge, TN 37831}

\author{Gary J.~Ferland}
 \affiliation{Department of Physics and Astronomy, University of Kentucky, Lexington, KY
 40506}

\date{\today}

\begin{abstract}
We present quantum mechanical close-coupling calculations of
collisions between two hydrogen molecules over a wide range of
energies, extending from the ultracold limit to the super-thermal
region. The two most recently published potential energy surfaces
for the H$_2$-H$_2$ complex, the so-called DJ (Diep and Johnson,
2000) and BMKP (Boothroyd et al., 2002) surfaces, are
quantitatively evaluated and compared through the investigation of
rotational transitions in H$_2$+H$_2$ collisions within rigid
rotor approximation. The BMKP surface is expected to be an
improvement, approaching chemical accuracy, over all conformations
of the potential energy surface compared to previous calculations
of H$_2$-H$_2$ interaction. We found significant differences in
rotational excitation/de-excitation cross sections computed on the
two surfaces in collisions between two para-H$_2$ molecules. The
discrepancy persists over a large range of energies from the
ultracold regime to thermal energies and occurs for several
low-lying initial rotational levels. Good agreement is found with
experiment (Mat\'e et al., 2005) for the lowest rotational
excitation process, but only with the use of the DJ potential.
Rate coefficients computed with the BMKP potential are an order of
magnitude smaller.
\end{abstract}

\maketitle


\section{INTRODUCTION}\label{intro}
Collisions involving two hydrogen molecules are of great interest
for three main reasons. First, the H$_2$+H$_2$ collision system is
a prototype for chemical dynamics studies and can be used as a
testing ground for scattering theory of inelastic (non-reactive)
diatom-diatom collisions involving a weak interaction potential.
Second, H$_2$ is the most abundant molecular species in the
universe. The rotational and vibrational transitions in H$_2$
induced by collisions with its twin are of practical importance in
models of astrophysical environments where the physical conditions
may not be accessible to terrestrial experiments. Examples include
low densities characteristic of giant molecular clouds in the
interstellar medium where star formation occurs and H$_2$ may act
as a coolant \cite{LeBoulot}. Heating of the interstellar cloud by
strong shock waves induces rotational and vibrational excitation
of the H$_2$ molecules leading to collision-induced dissociation
to two free H atoms \cite{Chang} and photodissociation regions
where H$_2$ is exposed to strong UV stellar radiation \cite{Shaw}.
Third, with recent experimental advances in the cooling and
trapping of molecules \cite{Doyle,Gould,Pillet,Knize} to form
molecular Bose-Einstein condensates,
collisional studies of the H$_2$+H$_2$ system can serve as a model
to provide new insights into the behavior of diatom-diatom
collisions at ultracold temperatures including investigations of
Feshbach resonances, predissociation in van der Waals complexes,
determination of complex scattering lengths, testing of effective
range theory and Wigner threshold laws, and quasiresonant
vibration-rotation energy transfer
\cite{Bala97,Bala98,Forrey98,Forrey99,Forrey01} in molecular
collisions.

Since (H$_2$)$_2$ is the simplest 4-identical-particle closed
shell system and there is a continuing demand for accurate
collisional data for modeling astrophysical, atmospheric and
combustion processes there have been considerable
theoretical\cite{Davison,Takayangi,Allison,Zarur,Green,
Heil,Schaefer,Dandy, Flower1,Flower2,Flower3,Flower4,Flower5}
 and experimental\cite{Dondi,Farrar,Audibert,Dove,Kreutz,Mate} studies
performed on the H$_2$+H$_2$ system. The most recent experiment
was performed by Mat\'{e} and co-workers \cite{Mate}. Using the
technique of Raman spectroscopy with supersonic expansions of
para-H$_2$, they measured the rate coefficient, $k_{00\rightarrow
20}$, for H$_2(j_1=0)$ + H$_2(j_2=0)\to$ H$_2(j_1'=2)$ +
H$_2(j_2'=0)$  collisions in the temperature range of 2 to 110 K.
With their experimental methodology, without the loss of
generality, the reduction of the measured $k_{00\rightarrow 20}$
rate coefficient to the corresponding cross section,
$\sigma_{00\rightarrow 20}$  in the collision energy range of  360
to  600 cm$^{–-1}$ was made possible. To complement the
experimental measurement, Mat\'{e} et al. \cite{Mate} also
performed
 coupled channel quantum  scattering calculations
to determine the $\sigma_{00\rightarrow20}$ excitation cross
section as a function of the collision energy. Good agreement was
found between the experimentally-derived cross sections and
theoretical results obtained using the rigid rotor potential
energy surface (PES) developed by Diep and Johnson (DJ)\cite{DJ}.

The most recent theoretical study of rotational excitation in
H$_2$+H$_2$ collisions  was carried out by Gatti and
co-workers\cite{Gatti}. They employed the  wave-packet propagation
method in conjunction with the multiconfiguration time-dependent
Hartree algorithm to compute rotational excitation cross sections
for collision energies up to 1.2 eV  by a flux analysis of the
interaction of the wave-packet with a complex absorbing potential.
Gatti et al. compared their  results with the wave packet
calculations of Lin and Guo \cite{LinGuo2002} who employed the
coupled-states (CS) approximation which neglects Coriolis
coupling. They found that the CS approximation can lead to rather
reliable predictions provided the calculations are performed at
low collision energies and low rotational excitations. However,
without the aid of accurate molecular potential curves, it is
difficult to determine in which energy regime the Coriolis
coupling takes effect for a specific collision system. Both
wave-packet calculations employed the
PES constructed in 2002 by Boothroyd, Martin, Keogh, and Peterson \cite{BMKP}  
(BMKP). While previous calculations\cite{Heil} have indicated
that the CS approximation can give accurate results for rotational
excitation in H$_2$ at low energies, these calculations have used
rather simple potential functions for the H$_4$ system.

The purpose of this study is to perform accurate close-coupling
calculations of rotational transitions in H$_2$+H$_2$ collisions
within the ground vibrational state. Due to the relatively large
vibrational level spacings of the H$_2$ molecule and the weak
H$_2$--H$_2$ interaction potential, the rigid rotor approximation
is expected to hold well for the present system for pure
rotational energy transfer.
 This is also confirmed by the
 wave packet calculations of Lin and Guo\cite{LinGuo2002}.
Here, we perform close-coupling calculations of rotational
excitation on both the DJ and the BMKP PESs and compare our
results with the wave-packet results of Lin and Guo and Gatti et
al. So far, the accuracy of the BMKP PES has not been established
by comparing results from accurate quantum calculations with
experimental results. Such a comparison is provided here for both
the DJ and BMKP PESs and we show that the rigid rotor DJ potential
yield results that are generally in better agreement with
experiment.

We also investigate the behavior of elastic and rotationally
inelastic collisions in diatom-diatom collisions at ultracold
energies by taking the H$_2$+H$_2$ system as an illustrative
example. While ultracold rotational and vibrational energy
transfer in atom-diatom collisions have extensively been
reported\cite{Bala97,Bala98,Bala00,Forrey98,Forrey99,Forrey01} in
the last several years, such calculations are yet to be performed
on molecule-molecule collisions.  Forrey\cite{Forrey01} has
performed limited calculations on H$_2$+H$_2$ collisions in the
ultracold regime using the semi-empirical potential of Zarur and
Rabitz\cite{Zarur} while Avdeenkov and
Bohn\cite{Avdeenkov2001,Avdeenkov2003,Avdeenkov2005}
 reported spin-exchange collisions in O$_2$+O$_2$ and OH+OH/OD+OD
systems. Here, we provide a detailed investigation of rotational
energy transfer in H$_2$+H$_2$ collisions on the DJ and BMKP PESs
at ultracold energies and report complex scattering lengths for
collisions involving different initial rotational levels of the
two molecules.

The paper is organized as follows: A brief description of the
theoretical methodology is given in section \ref{calc} and results
are presented in section \ref{results}. Section \ref{summary}
provides summary and conclusions. Atomic units are used
throughout, unless otherwise noted: i.e.,
$\hbar=e=m_e=a_o=1$~a.u., while 1 hartree = 27.2116 eV = 627.51
kcal/mol.

\section{THEORY}\label{calc}

Calculations of state-to-state rotational transition cross
sections and rate coefficients can provide an important test of
the reliability of the potential energy surfaces describing the
interaction of two H$_2$ molecules when compared to available
experiments. To compute the scattering amplitudes and hence the
cross sections, we use well established quantum mechanical
close-coupling methods \cite{Takayangi,Green,Secrest,Child}. Here
we only summarize the essence of the theory. To describe the
scattering process, we solve the time-independent Schr\"{o}dinger
equation within rigid rotor approximation for the collision of two
H$_2$ molecules in the center of mass frame given by
\begin{equation}
\left(\hat{T}(R)+\sum^2_{i=1}\hat{h}_{rot}(\hat{r}_i)
 +V(\vec{R},\vec{r}_1,\vec{r}_2)-E\right)
  \Psi(\vec{R},\hat{r}_1,\hat{r}_2)=0,
  \label{FTISE}
\end{equation}
where $\hat{T}(R)$=$-\frac{1}{2\mu}\nabla^2_R$ is the kinetic     
energy operator and $\hat{h}_{rot}(\hat{r}_i)=
\frac{\hat{j}_{i}}{2\mu_{i}r^2_i}$ is the diatom rotational
kinetic energy operator; $\mu$ and $\mu_{i}$ are the reduced
masses of the H$_2$
collision pair and a isolated H$_2$ molecule, respectively. The internuclear  
distance between the two H atoms is denoted by $r_i$, and $R$ is
the distance between the center of mass of the diatoms;
$\hat{r}_1$ and $\hat{r}_2$ are the orientation angles of the
rotors 1 and 2, respectively. The term
$V$($\vec{R},\vec{r}_1,\vec{r}_2$) represents the H$_2$--H$_2$
interaction potential.

The rotational part of the Hamiltonian operator satisfies the
eigenvalue equation
\begin{equation}
\left(
\hat{h}_{rot}(\hat{r}_i)-B_ij_i(j_i+1)\right)Y_{j_im_i}(\hat{r}_i)=0
\end{equation}
where $B_i$ are the rotational constants of the rigid rotors.

The rotational angular momenta of the two molecules $\vec{j}_1$
and $\vec{j}_2$ are coupled to form $\vec{j}_{12}$, which is
subsequently coupled to the orbital momentum $\vec{l}$ to yield
the total angular momentum $\vec{J}$. The angular wave function in
the total angular momentum representation is given by
\begin{eqnarray}
 \phi^{JM}_{j_1 j_2 j_{12} l}(\hat{R},\hat{r}_1,\hat{r}_2)&=&
  \sum_{m_1 m_2 m_{12}m_l}
   (j_1m_1j_2m_2|j_{12}m_{12})
    (j_{12}m_{12}lm_l|JM) \nonumber \\
    && \times
     Y_{j_1m_1}(\hat{r}_1)Y_{j_2m_2}(\hat{r}_2)Y_{lm_l}(\hat{R})
\end{eqnarray}
and under spatial inversion
\begin{equation}
\mathcal{P}\phi^{JM}_{\alpha l}(\hat{R},\hat{r}_1,\hat{r}_2)=
(-1)^{j_1+j_2+j_{12}}\phi^{JM}_{\alpha
l}(\hat{R},\hat{r}_1,\hat{r}_2)
\end{equation}
where $\alpha\equiv j_1 j_2 j_{12}$, and
$m_1$, $m_2$, $m_{12}$ and $m_l$ are  the projections of    
$\vec{j}_1$, $\vec{j}_2$, $\vec{j}_{12}$ and $\vec{l}$,
respectively, onto the space-fixed $z$-axis.                
The symbol $(j_1j_2m_1m_2|JM)$ denotes a Clebsch-Gordon
coefficient.

Using the basis set expansion method, we expand our total wave
function ansatz as
\begin{equation}
\Psi(\vec{R},\hat{r}_1, \hat{r}_2)=
 \frac{1}{R} \sum_{JM\alpha l}
  F^{JM}_{\alpha l}(R) \phi^{JM}_{\alpha l}
   (\hat{R},\hat{r}_1,\hat{r}_2)
   \label{totalwf}
\end{equation}
 Substituting eqn.(\ref{totalwf}) into
eqn.(\ref{FTISE}), we arrive at a system of close-coupling
equations
\begin{equation}
\left( \frac{d^2}{dR^2}-\frac{l(l+1)}{R^2}+2\mu
 E_k\right)F^{J}_{\alpha l}(R) \\
 = 2\mu\sum_{\alpha' l'}
F^{J}_{\alpha' l'}(R)\langle\phi^{JM}_{\alpha
l}|V|\phi^{JM}_{\alpha' l'}\rangle, \label{cceqn}
\end{equation}
where the quantities in $\langle\cdots\rangle$ are the matrix
elements of the interaction potential and $E_k
=E-B_1j_1(j_1+1)-B_2j_2(j_2+1)$ is the kinetic energy of the
relative motion for a given value of the total energy $E$. The
solution of the coupled equations and asymptotic analysis of the
radial wave functions yield the scattering S-matrix from which
cross sections for state-to-state rotational transitions
 from an initial level specified by quantum numbers $j_1j_2$
to final levels $j_1'j_2'$  are given by
\begin{equation}
\sigma_{j_1j_2 \rightarrow j_1' j_2'}(E_k)=
 \frac{\pi}{2\mu E_k(2j_1+1)(2j_2+1)}
  \sum_{Jj_{12}j_{12}'ll'}(2J+1)
   |\delta_{\nu \nu'}-\mathcal{S}^{J}_{\nu \nu'}|^2,
\end{equation}
where $\nu\equiv j_1 j_2 j_{12} l$.

The above expression for cross section assumes that the two
diatomic molecules are distinguishable. However, for H$_2$-H$_2$
collisions the target and projectile molecules are
indistinguishable and one must take into account the symmetry of
the wave function under exchange. Thus, properly symmetrized total
angular momentum wave functions\cite{Green}
\begin{eqnarray}
 \phi^{JM\pm}_{j_1 j_2 j_{12} l}(\hat{R},\hat{r}_1,\hat{r}_2)&=&
\frac{1}{\sqrt{2(1+\delta_{j_1j_2})}}\left[ \phi^{JM}_{j_1 j_2
j_{12} l}(\hat{R},\hat{r}_1,\hat{r}_2)\right.
\nonumber \\
&& \pm \left.(-1)^{j_1+j_2+j_{12}+l}\phi^{JM}_{j_1 j_2 j_{12}
l}(\hat{R},\hat{r}_1,\hat{r}_2)\right]
\end{eqnarray}
need to be employed in which the index pair $j_1j_2$ is restricted
to $j_1\geq j_2$ to obtain a linearly independent set. Using the
symmetrized angular wave functions, one obtains coupled equations
similar to eqn. (\ref{cceqn})  which yield scattering cross
sections\cite{Green}

\begin{equation}
\sigma_{j_1j_2 \rightarrow j_1' j_2'}(E_k)=
 \frac{\pi (1+\delta_{j_1j_2}) (1+\delta_{j_1'j_2'})}{2\mu E_k(2j_1+1)(2j_2+1)}
  \sum_{Jj_{12}j_{12}'ll'}(2J+1)
   |\delta_{\nu \nu'}-\mathcal{S}^{J}_{\nu \nu'}|^2.
\label{indist-cxn}
\end{equation}

Rate coefficients for state-to-state rotational transitions are
obtained by averaging the appropriate cross sections over a
Boltzmann distribution of  relative speeds  of the projectile
molecule at a given temperature $T$:
\begin{equation}
k_{j_1 j_2 \rightarrow j_1' j_2'}(T)= G
\int_{0}^{\infty}dE_k\sigma_{j_1 j_2 \rightarrow j_1'
j_2'}(E_k)E_k e^{(-\beta E_k)},
\end{equation}
where the constant $G = \sqrt{\frac{8}{\mu \pi \beta}}\beta^2$ and
$\beta =(k_B T)^{-1}$ with $k_B$ being the Boltzmann constant. The
total quenching rate coefficient can be calculated from
\begin{equation}
k_{j_1 j_2}(T)= \sum_{j_1' j_2'} k_{j_1 j_2 \rightarrow j_1'
j_2'}(T).
\end{equation}

\section{RESULTS}\label{results}

We have carried out close-coupling calculations for collisions of
H$_2$ with H$_2$ using the BMKP and DJ PESs. The rigid rotor
target and projectile energy levels were calculated using a
rotational constant of {\it B} = 60.853 cm$^{-1}$ for the H$_2$
molecule.
 To solve the coupled radial equations (\ref{cceqn}), we
used the hybrid modified log-derivative-Airy propagator
\cite{Manolopoulos} in the general purpose non-reactive scattering
code MOLSCAT \cite{Molscat}. The log-derivative matrix
\cite{Manolopoulos} is propagated to large intermolecular
separations where the numerical results are matched to the known
asymptotic solutions to extract the physical scattering matrix.
This procedure is carried out for each partial wave until a
converged cross section is reached. We have checked that the
results are converged with respect to the number of partial waves
as well as the matching radius for all channels included in the
calculations.

In addition to the partial wave convergence, based on the DJ PES,
we have checked that the results are converged with respect to
various parameters that enter into the close-coupling
calculations. These include the number of quadrature points used
for angular integration, the number of terms in the angular
expansion of interaction potential, and the asymptotic matching
radius for radial integration. In the ultracold regime, we used a
matching radius of $R_o$ = 200--300 a.u. to obtain converged
values of elastic and inelastic cross sections   while for the
subthermal energy region, a matching radius of 50 a.u. was
sufficient to yield results of comparable accuracy. Similarly, we
used 10 quadrature points each for integration along angular
coordinates $\theta_1,~\theta_2$, and $\phi_{12}$.

Finally, two different basis sets (22-state: $j_1$$j_2$ = 00, 20,
22, $\cdot \cdot \cdot$, 44 and 50-state: $j_1$$j_2$ = 00, 20,
22,$\cdot\cdot\cdot$, 66) were also employed to further test the
convergence of our results. For  $E<$ 1.0 eV, the two basis sets
yield results within 1\%, and
 at $E$ = 1.0 eV, a similar
degree of accuracy was obtained for the dominant transitions. For
weaker transitions such as 00 $\rightarrow$ 44 rotational
excitation, the small and large basis sets gave
$\sigma_{00\rightarrow 44}$ = 3.01$\times$10$^{-18}$  and
 3.20$\times$10$^{-18}$ cm$^{2}$, respectively, at 1.0 eV. Note that
the cross sections for dominant transitions are two orders of
magnitude larger than the weaker ones. Even at $E$ = 2.6 eV, the
cross sections obtained from both basis sets for the dominant
transitions have similar convergence properties as for $E$ = 1.0
eV. However, for 00$\rightarrow$44, we found that the larger basis
set is preferred since there is a significant difference between
the cross sections (i.e., $\sigma_{00\rightarrow 44}$ =
4.31$\times$10$^{-17}$ cm$^{2}$ and 6.30$\times$10$^{-17}$
cm$^{2}$, respectively,  from the small and large basis sets, at
2.6 eV). Since our focus is on the low-energy region where the two
basis sets yield similar results, the smaller basis set is adopted
throughout the calculations.

Fig.~1 shows the comparison between theoretical and experimental
rate coefficients for the 00$\rightarrow$20 transition in the
temperature range between 50 K and 300 K. Both experimental and
theoretical rate coefficients indicate a precipitous drop for
temperatures lower than 100 K. Unexpectedly, we find that the
00$\rightarrow$20 excitation rate coefficient computed with the
BMKP PES is about an order of magnitude smaller than that
calculated with the PES of DJ and the experiment, though both BMKP
and DJ display the same trend. Only  results from the DJ PES agree
with the experimental data of Mat\'{e} {\it et al.} \cite{Mate}
Good agreement between experiment and theory based on the DJ PES
was also shown by Mat\'{e} {\it et al.} \cite{Mate}. Further, the
theoretical results obtained by Flower \cite{Flower1} using the
older PES of Schwenke \cite{Schwenke} are also seen to be in good
agreement with experiment. The discrepancy with the BMKP results
may directly be traced to the weaker anisotropy of the BMKP PES
responsible for the 00$\rightarrow$20 transition.
In the calculations the angular dependence of the interaction
potential is represented as \cite{DJ}
\begin{equation}
V(R,\theta_1, \theta_2, \phi_{12}) = \sum_{l_1,l_2,l}
V_{l_1,l_2,l}(R)G_{l_1,l_2,l}(\theta_1, \theta_2, \phi_{12}),
\end{equation}
where $V_{l_1,l_2,l}(R)$ are radial expansion coefficients and
$G_{l_1,l_2,l}$($\theta_1$, $\theta_2$, $\phi_{12}$) are
bispherical harmonics. In Fig.\ 2 we compare the spherically
symmetric ($V_{000}$) and the leading anisotropic terms,
$V_{022}=V_{202}$ and $V_{224}$, in the angular expansion of the
BMKP and DJ interaction potentials as functions of the
intermolecular separation. It is seen that while the spherically
symmetric part is nearly identical for both potentials, the main
anisotropic term, $V_{022}=V_{202}$,
 responsible for the 00$\rightarrow$20 rotational excitation is smaller for the BMKP potential
at small intermolecular separations. Our test calculations show
that at energies lower than 0.1 eV, the discrepancy between the
two results is mostly due to small differences in the coupling
elements in the region of the van der Waals minimun, i.e., $R>3.0$
au. The same also applies to the  next higher order term,
$V_{224}$. On the other hand,
  Progrebnya and Clary\cite{clary} found that the BMKP surface yields
too high values for vibrational relaxation in
H$_2(v=1)$+H$_2(v=0)$ collisions. They attributed this to
higher-order anisotropic terms in the BMKP PES that leads to
preferential population of high rotational levels in the $v'=0$
level after quenching (see Fig.\ 2 of the above reference).
 Calculations employing a modified
version of the BMKP potential in which only the first two leading
anisotropic terms of the interaction potential shown in Fig.\ 2
are retained gave results in better agreement with the experiment.

The integral elastic cross sections of para-H$_2$+para-H$_2$
collisions as a function of collision energy is plotted in
Fig.~3(a) for both the BMKP and DJ potentials. In the zero-energy
limit, the elastic cross sections attain finite values in
accordance with Wigner's law. The limiting value of the elastic
cross section is $1.91\times 10^{-13}$ cm$^2$ and $1.74\times
10^{-13}$ cm$^2$ for the  BMKP and DJ PESs, respectively.
 The comparable values of
the limiting elastic cross sections on the two potentials is
explained based on the nearly identical values for the spherically
symmetric part of the interaction potentials for both surfaces
(see Fig.\ 2). Both potentials exhibit a shape resonance at
collision energies between 0.0002 and 0.0003 eV, arising from the
$l=2$ partial wave. At higher energies, cross sections on the two
potentials exhibit
an oscillatory behaviour (see inset in the top panel of Fig.\ 3)      
 which arises from interference between
partial cross sections corresponding to different values of the
total angular momentum quantum number $J$. Note that only even
values of $J$ are allowed for p-H$_2$--p-H$_2$ collisions.
Schaefer and Meyer\cite{Schaefer} have provided a detailed
analysis of the oscillatory behavior of the elastic cross
sections.

In Fig.~3(b) we compare elastic cross sections from the present
work on the DJ and BMKP surfaces with the theoretical results of
Diep and Johnson\cite{DJ} and the experimental measurement of
Bauer et al.\cite{Bauer}. It is seen that the results on the DJ
potential give slightly better agreement with experimental data
although the overall agreement between the experiment and theory
is generally good.

In Fig.~4 we compare cross sections from the present calculations
on the BMKP surface for the $00\rightarrow20$, $00\rightarrow22$
and $00\rightarrow40$ transitions  with the wave packet results of
Gatti et al.\cite{Gatti} and Lin and Guo \cite{LinGuo2002}.
 The overall agreement between the CC and the wave-packet
results is rather good, to within 10--15\% for all cross sections,
except for the $00\rightarrow22$ rotational excitation for which
the present results are 30--50\% larger. The significant
differences between the present results  and the wave packet
results for this transition is somewhat surprising considering the
fact that the corresponding cross sections are about an order of
magnitude larger than the $00\rightarrow40$ transition for which
we obtain good agreement with the wave packet results. We are
confident that this is not a numerical error in our calculations
as  we have  benchmarked our results against a new
time-independent coupled channel code developed by
Krems\cite{krems} which  reproduces our results up to several
significant digits for all
transitions shown in Fig.\ 4.                  
The deviation of the CS wave-packet results of  Lin and Guo from
the present CC and full wave-packet data of Gatti et al. at higher
energies may be attributed to Coriolis couplings. This evidently
suggests that Coriolis coupling plays an important role at higher
energies and that the rigid rotor approximation appears to hold
well for this system for collision energies investigated in the
present work.

In Fig.\ 5(a), we compare rotational quenching cross sections for
the $20\rightarrow00$ transition evaluated using the BMKP and DJ
potential with the $20\rightarrow20$ elastic scattering cross
section. Since the elastic scattering cross section on the two
PESs is comparable, only the result on the DJ potential is shown.
 The resonance feature that occurs in all three cross sections just above 10$^{-4}$
eV is due the $l=2$ shape resonance  discussed previously (See
Fig.\ 1). It is seen that the inelastic cross sections are much
smaller than the elastic one at all energies shown in the Fig.\
5(a). This suggests that it may be possible to cool rotationally
excited H$_2$ molecules in the $j=2$ rotational level by
thermalizing collisions with ground state H$_2$ molecules
(evaporative cooling) without significant trap loss, though the
absence of an electric dipole moment makes it a difficult system
to handle experimentally. For incident  energies lower than
$10^{-5}$ eV, the quenching cross section varies inversely with
the velocity in accordance with Wigner threshold behavior. As a
consequence, the product of the relative velocity and the
quenching cross section attains a finite value in the limit of
zero incident kinetic energy, as illustrated in Fig.\ 5(b). The
limiting value of the quenching rate coefficient is $2.4\times
10^{-13}$ cm$^3$/s.

In ultracold collisions where $s$-wave scattering dominates,
elastic and inelastic scattering cross sections are conveniently
expressed in terms of scattering lengths. The scattering length is
real when only elastic scattering is present, but it becomes
complex with the inclusion of inelastic channels\cite{Bala98}. The
complex scattering length is given by
$a_{j_1j_2}=\alpha_{j_1j_2}-i\beta_{j_1j_2}$ where $\alpha$ and
$\beta$ are real and imaginary parts of the scattering length. The
imaginary part of the scattering length $\beta$ is related to the
zero-temperature limit of the quenching rate coefficient:
$k_{j_1j_2}(T\to 0)=4\pi\beta_{j_1j_2}\hbar/\mu$. For the DJ
potential we obtain the values $\alpha_{00}=5.88$~\AA,
$\alpha_{20}=5.78$~\AA, $\beta_{20}=0.003$~\AA,
$\alpha_{22}=5.83$~\AA, and $\beta_{22}=0.0023$~\AA. The
corresponding values for scattering on the BMKP potential are
$\alpha_{00}=6.16$~\AA, $\alpha_{20}=6.16$~\AA,
$\beta_{20}=0.00028$~\AA, $\alpha_{22}=6.16$~\AA, and
$\beta_{22}=0.00071$~\AA. It is seen that the real part of the
scattering length remains practically unchanged for the three
initial states indicating that no zero-energy resonances occur for
any of the three initial states on either PES. The presence of
zero-energy resonances (bound/quasibound states near channel
thresholds) generally enhances the elastic scattering cross
section at low energies. The smaller value of the inelastic
quenching rates on the BMKP surface is also reflected in the
values of $\beta$ for all three initial states.

In Fig.\ 6 we compare cross sections for $22\rightarrow00$ and
$22\rightarrow20$ transitions obtained using the DJ potential with
the CC calculations of Forrey \cite{Forrey01} based on the PES of
Zarur  and Rabitz \cite{Zarur}. The agreement is remarkably good
considering that the DJ potential is derived from accurate ab
initio calculations while that of Zarur  and Rabitz is a model
semi-empirical potential. The reproduction of the shape resonance
near $E=10^{-4}$ eV by the two potentials is a good indicator of
the accuracy of the two potential surfaces.
 The
solid curve in Fig.\ 6 is the quenching cross section for
$20\rightarrow00$ transition on the DJ potential and it is seen
that the cross sections for $22\rightarrow 20$ transition is
larger at all energies.


Further comparison between BMKP and the DJ potential is presented
in Fig.\ 7 in which we provide energy dependence of the excitation
cross sections for $00\rightarrow20$, 22, 40, 42 and 44
transitions. The general trend in all cases, except for
$00\rightarrow 40$ at low energies, is that the BMKP potential
yields smaller values of cross sections compared to the DJ
potential. The differences get somewhat smeared out when the cross
sections are integrated over a Boltzmann distribution of relative
velocities of the two molecules to yield the rate constants. This
is  illustrated in Fig.\ 8 for the same transitions as given in
Fig.\ 7.  Comparison is also made to the rate coefficients
computed by Flower \cite{Flower1} which show better agreement with
the DJ results.

\section{SUMMARY AND CONCLUSIONS}\label{summary}

We have performed  quantum close-coupling calculations of elastic
and inelastic rotational transitions in collisions of H$_2$ with
H$_2$ using  the two most recently published {\it ab initio}
interaction potential energy surfaces (PESs) for the H$_4$ system.
The calculations span a wide range of energies (9-orders of
magnitude)
 extending from the zero-temperature
limit to about 2.0 eV. Sensitivity of the results to details of
the interaction potential is presented by computing real and
imaginary parts of the scattering lengths for different initial
rotational levels of the two colliding H$_2$ molecules. It is
shown that the limiting elastic cross section is not very
sensitive to the initial rotational levels of the two H$_2$
molecules although the inelastic cross sections strongly depend on
the initial rotational level.

We also showed that results obtained using the rigid rotor
potential surface of
  Diep and
Johnson\cite{DJ} are in close agreement with the experimental
measurements of  Mat\'{e} {\it et                              
al.} \cite{Mate} for $00 \rightarrow 20$ rotational excitation
rate coefficient.
 However, the corresponding results
 obtained using the Boothroyd {\it et al.}
\cite{BMKP} PES are a factor of ten smaller. This is quite
significant because the Boothroyd et al. surface is generally
believed to be the most accurate potential surface for the H$_4$
system and it has been used in two most recent six-dimensional
quantum mechanical calculations \cite{LinGuo2002,Gatti} of
rotational excitation in H$_2$+H$_2$ collisions.
The present study demonstrates that the BMKP surface will need to
be reevaluated before it can be adopted in large-scale scattering
calculations, especially for astrophysical applications.



\section{ACKNOWLEDGMENTS}
TGL and GJF acknowledges support from NASA grant NNG05GD81G and
the Spitzer Space Telescope Theoretical Research Program. The work
of RCF was supported by NSF grants PHY-0244066 and PHY-0554794. NB
acknowledges support from NSF grant PHY-0555565 and DOE grant
DE-FG36-05GO85028. PCS acknowledges support from NSF grant
AST-0087172. We acknowledge support from the Institute for
Theoretical Atomic, Molecular, and Optical Physics at the
Harvard-Smithsonian Center for Astrophysics for a workshop which
initiated this work.

\newpage

\begin{figure}[h]
\includegraphics[width=\figwidth]{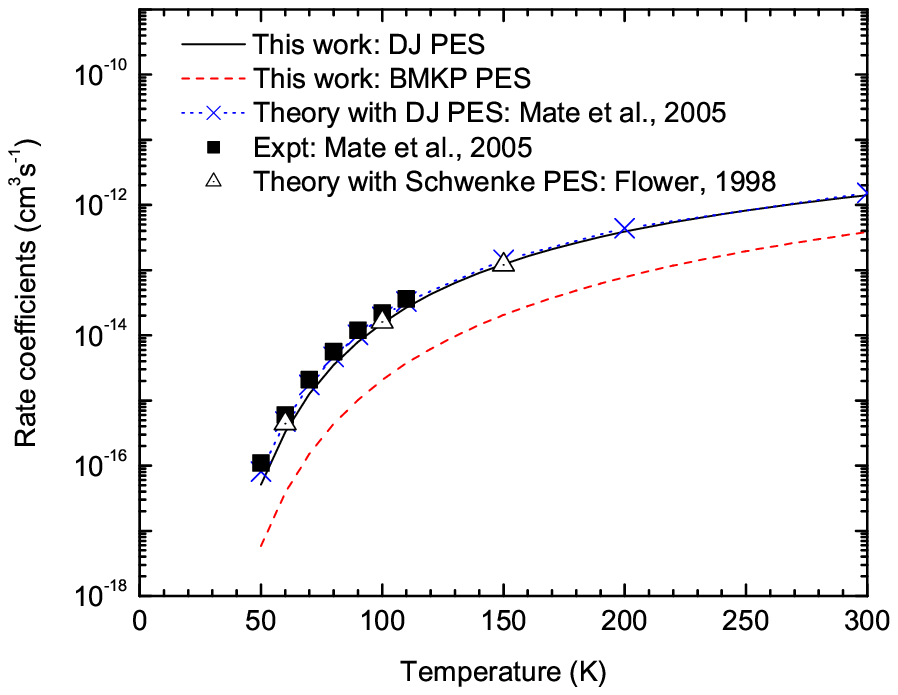}
\caption{Rate coefficients for $00\rightarrow20$  rotational excitation
in H$_2$+H$_2$ collisions as a function of temperature. The solid and dashed curves are
results obtained using the DJ and BMKP potentials, respectively.
The solid squares are the experimental results and the
dotted line with crosses is the theoretical calculation based on DJ PES
by Mat\'{e} et al.\cite{Mate}. The results on the DJ potential agree very well with
the experiment data whereas those obtained with BMKP do not.}
\end{figure}

\begin{figure}[h]
\includegraphics[width=\figwidth]{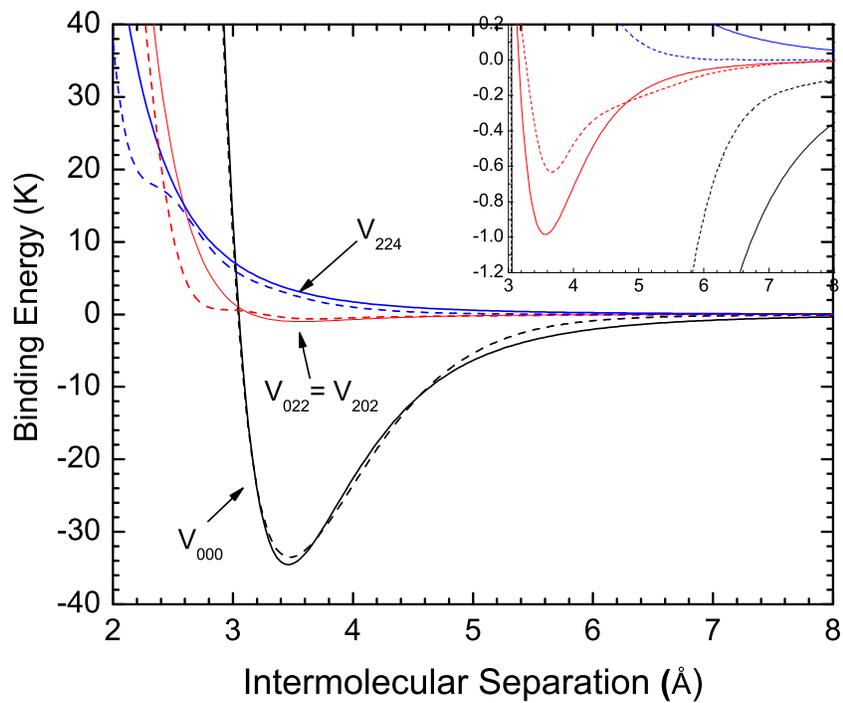}  
\caption{First three terms in the angular expansion of the intermolecular
potential as functions of the intermolecular  radial separation. The solid curves represent the
DJ potential and the dashed curves denote the BMKP potential.}
\end{figure}

\begin{figure}[h]
\includegraphics[width=\figwidth]{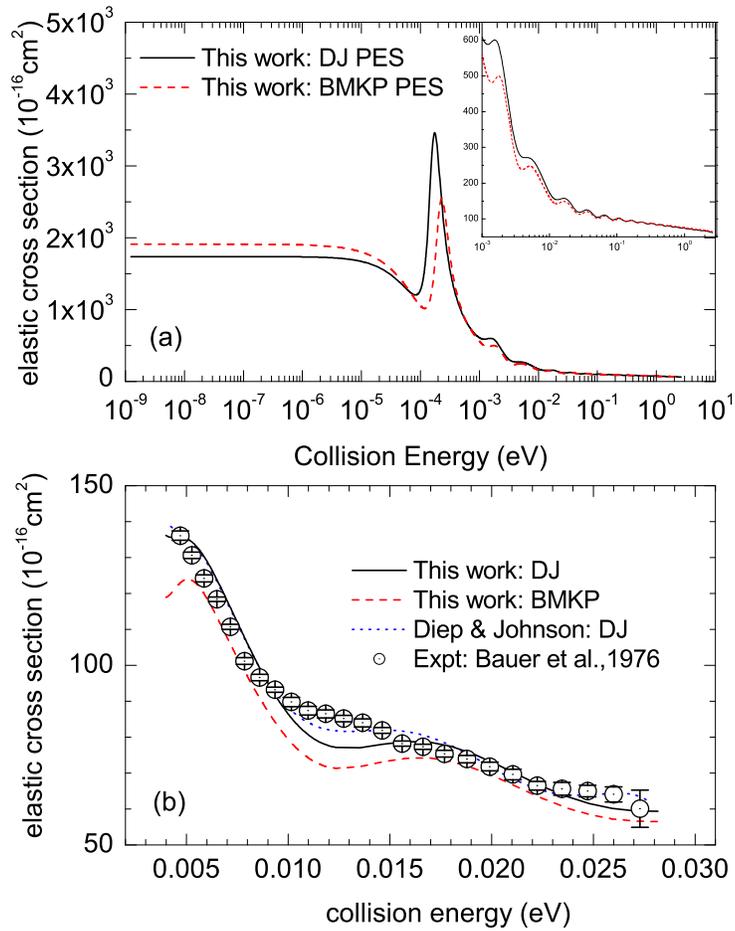}
\caption{Elastic cross section $\sigma_{00\rightarrow 00}$ as a
function of collision energy. The solid and dashed curves represent
the results on the DJ and BMKP surfaces, respectively. The dotted line is from Diep and
Johnson\cite{DJ}. The circles with error bars are
measurements of Bauer et al.\cite{Bauer}.}
\end{figure}

\begin{figure}[h]
\includegraphics[width=\figwidth]{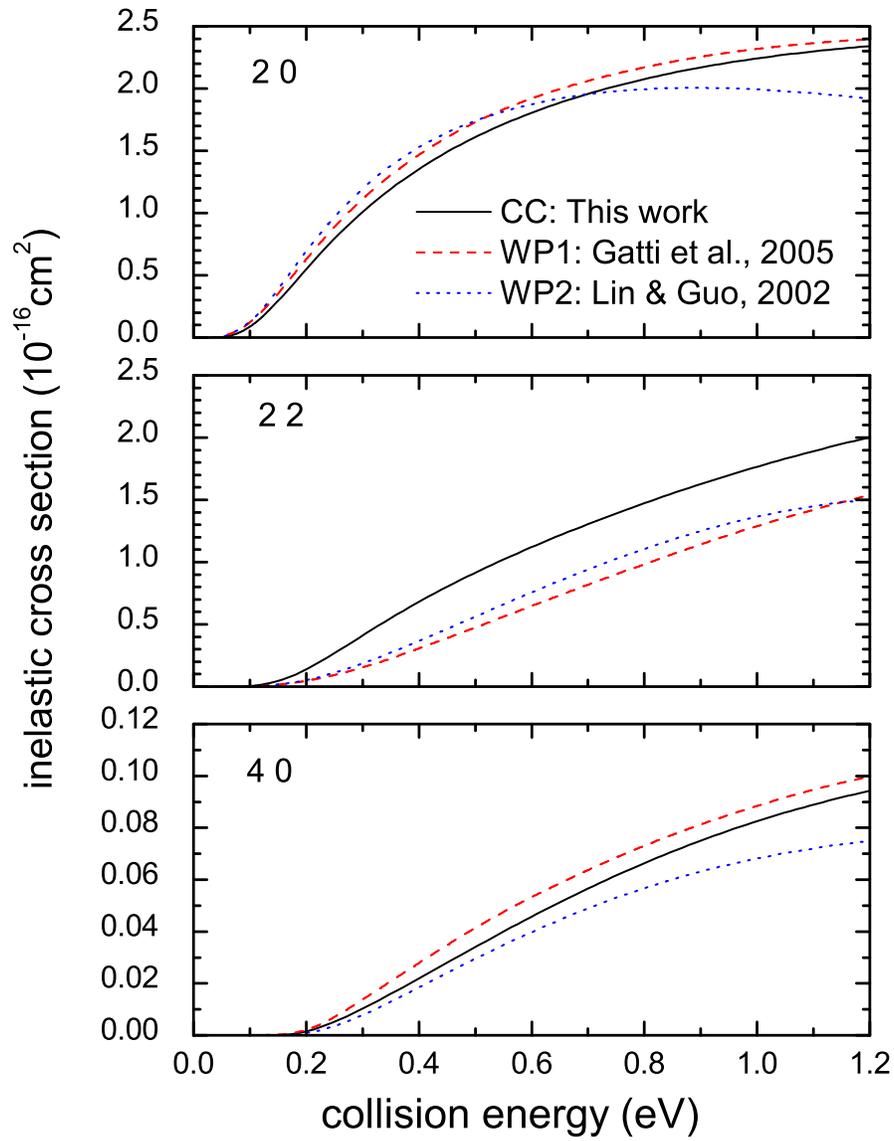}
\caption{Comparison of cross sections for $00\rightarrow20$, 22 and
40 transitions. The calculations are based on the BMKP PES. The solid,
dashed and dotted lines are the present close-coupling calculation,
wave-packet calculation of Gatti et al.\cite{Gatti} and
Lin and Guo\cite{LinGuo2002}, respectively.}
\end{figure}

\begin{figure}[h]
\includegraphics[width=\figwidth]{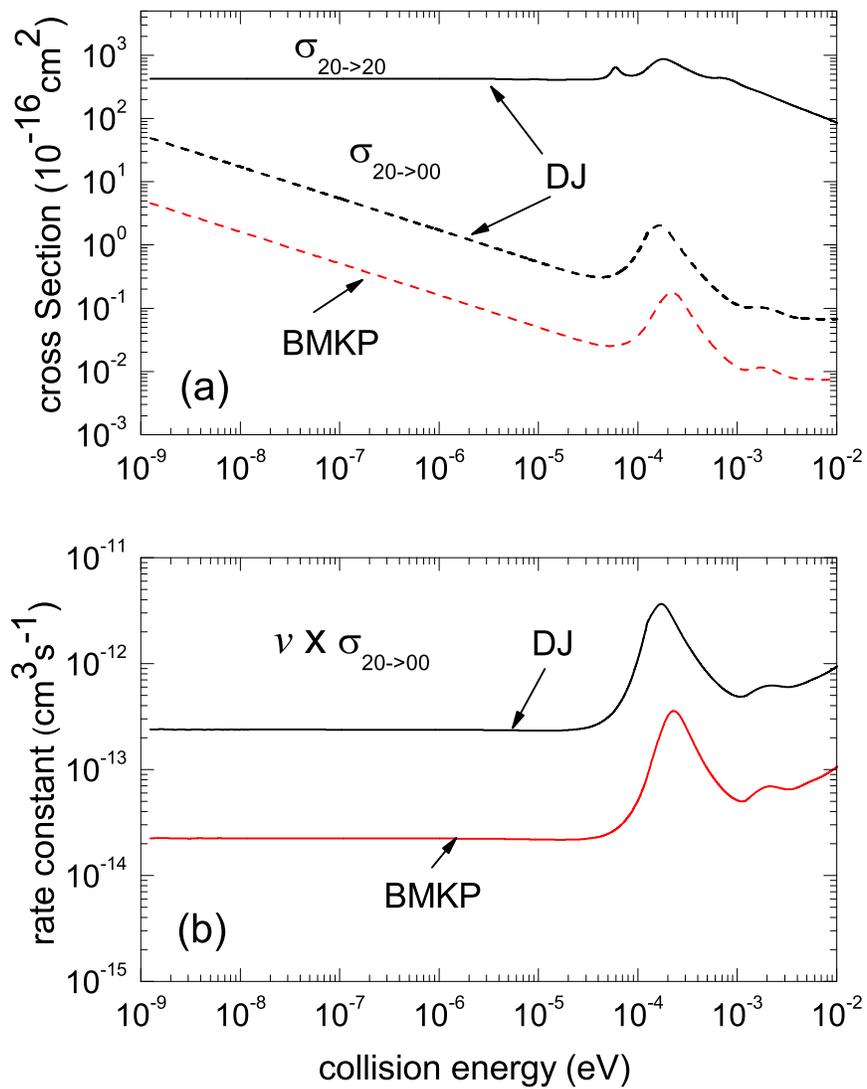}
\caption{Cross sections for $20\rightarrow00$ rotational quenching as a
function of collision energy. (a) Comparison with $20\rightarrow20$
elastic cross section. (b) Energy dependent rate coefficients, i.e.,
relative velocity times cross section.}
\end{figure}

\begin{figure}[h]
\includegraphics[width=\figwidth]{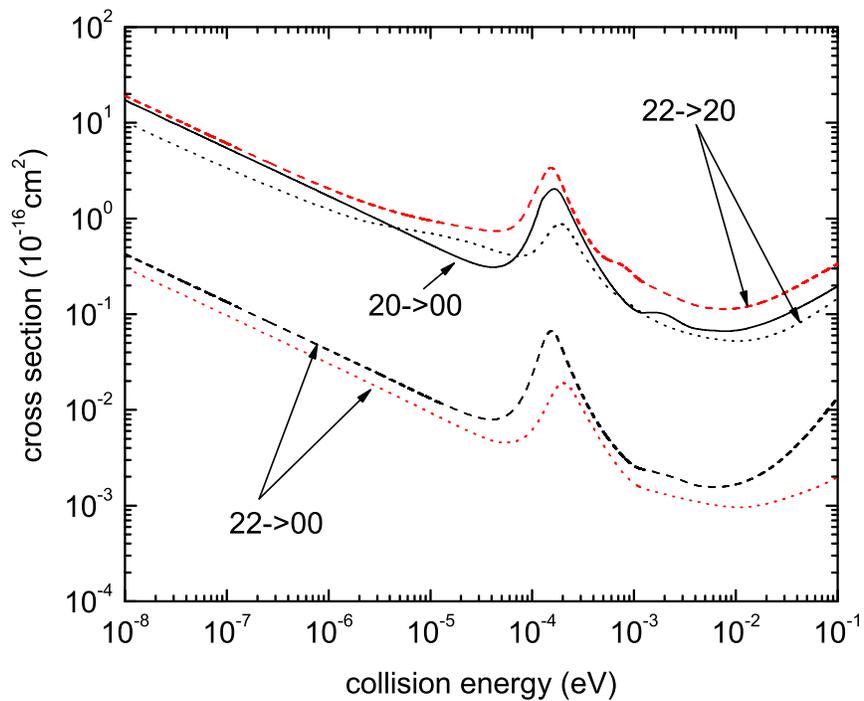}
\caption{Cross sections for $22\rightarrow00$ and 20 quenching collisions
as functions of collision energy plotted against
$20\rightarrow00$ cross sections. The calculations are based on DJ
PES. The dotted lines are the results of Forrey\cite{Forrey01}.}
\end{figure}

\begin{figure}[h]
\includegraphics[width=\figwidth]{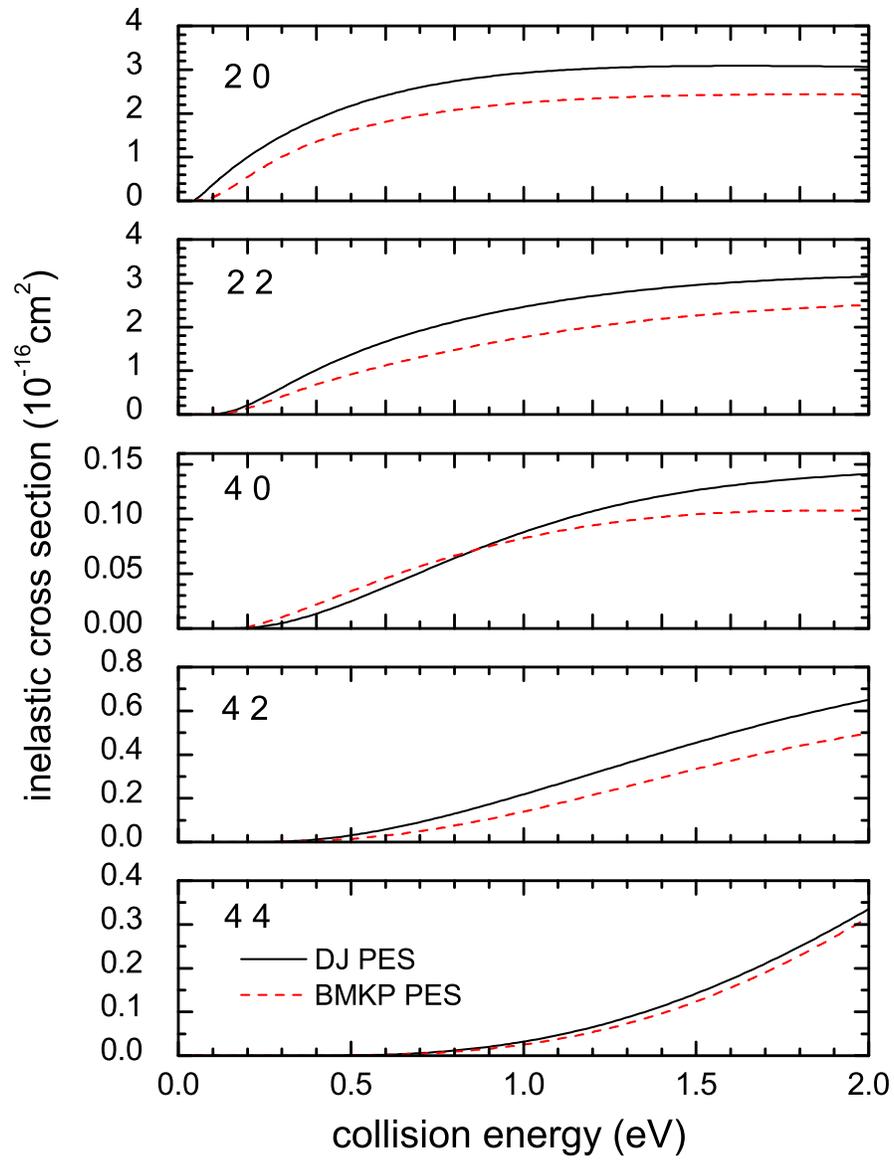}
\caption{Comparison of the cross sections for
$00\rightarrow20$, 22, 40, 42 and 44 transitions as functions of the
collision energy. The solid and dashed lines
represent DJ and BMKP PESs, respectively.}
\end{figure}

\begin{figure}[h]
\includegraphics[width=\figwidth]{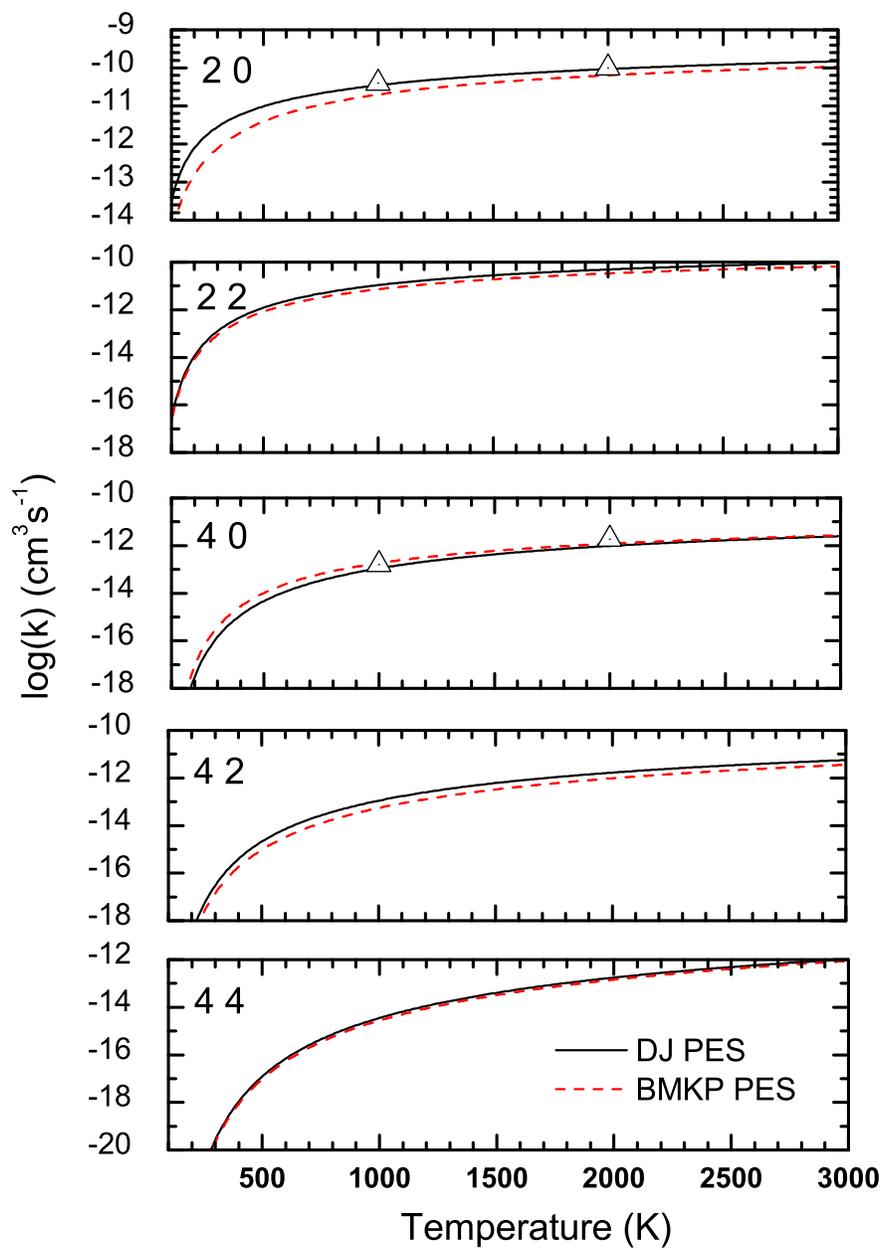}
\caption{Similar to Fig.\ 7, except for the rate coefficient as a
function of the temperature.}
\end{figure}

\end{document}